\documentclass[runningheads]{llncs}
\usepackage[T1]{fontenc}
\usepackage{float}
\usepackage{graphicx}
\usepackage{authblk}
\usepackage{amssymb}
\usepackage{marvosym}
\usepackage{xcolor}
\usepackage{booktabs} 
\usepackage{subcaption} 
\usepackage{colortbl} 

\newcommand\blfootnote[1]{
    \begingroup
    \renewcommand\thefootnote{}\footnote{#1}
    \addtocounter{footnote}{-1}
    \endgroup
}

\begin{document}

%
%

\title{Uncovering cognitive taskonomy through transfer learning in masked autoencoder-based fMRI reconstruction}
\titlerunning{Uncovering cognitive taskonomy through transfer learning}
\author{
Youzhi Qu\inst{1}\textsuperscript{,*} \and
Junfeng Xia\inst{1}\textsuperscript{,*} \and
Xinyao Jian\inst{2} \and
Wendu Li\inst{1} \and
Kaining Peng\inst{1} \and
Zhichao Liang\inst{1} \and
Haiyan Wu\inst{3} \and
Quanying Liu\inst{1}\textsuperscript{(\Letter)} 
}

%
\authorrunning{Y. Qu and J. Xia et al.}
%
\institute{
Department of Biomedical Engineering, Southern University of Science and Technology, Shenzhen, China\\
\and
Department of Biostatistics, Epidemiology and Informatics, Perelman School of Medicine, The University of Pennsylvania, Pennsylvania, USA\and
Centre for Cognitive and Brain Sciences and Department of Psychology, University of Macau, Macau, China \\
\email{liuqy@sustech.edu.cn}\\
}

\maketitle              
\begin{abstract}
Data reconstruction is a widely used pre-training task to learn the generalized features for many downstream tasks. Although reconstruction tasks have been applied to neural signal completion and denoising, neural signal reconstruction is less studied. Here, we employ the masked autoencoder (MAE) model to reconstruct functional magnetic resonance imaging (fMRI) data, and utilize a transfer learning framework to obtain the \textit{cognitive taskonomy}, a matrix to quantify the similarity between cognitive tasks. Our experimental results demonstrate that the MAE model effectively captures the temporal dynamics patterns and interactions within the brain regions, enabling robust cross-subject fMRI signal reconstruction. The cognitive taskonomy derived from the transfer learning framework reveals the relationships among cognitive tasks, highlighting subtask correlations within motor tasks and similarities between emotion, social, and gambling tasks. Our study suggests that the fMRI reconstruction with MAE model can uncover the latent representation and the obtained taskonomy offers guidance for selecting source tasks in neural decoding tasks for improving the decoding performance on target tasks.

%
%
%

\keywords{FMRI reconstruction  \and Transfer learning \and Masked autoencoder \and Cognitive tasks \and Taskonomy.}
\end{abstract}

\blfootnote{\textsuperscript{*} These authors contributed equally to this work.}

\section{Introduction}

In computer vision and natural language processing, reconstruction tasks serve as effective pre-training methods, guiding models to learn generalized features and achieving favorable performance on downstream tasks~\cite{devlin2019bert,he2022masked}.
In computer vision, reconstruction tasks are widely applied in image inpainting, denoising, and super-resolution~\cite{tian2020deep,wang2020deep,xiang2023deep}. In natural language processing, reconstruction tasks can improve the accuracy and fluency of translation, and enhance the completeness and creativity of content generation~\cite{brown2020language,bubeck2023sparks}. As pre-training tasks, reconstruction tasks enhance the models' ability to represent data by training them to reconstruct missing information. This capability not only enhances model performance on reconstruction tasks but also boosts performance on downstream tasks such as image classification, semantic understanding, and sentiment analysis. 
Classification tasks within neural decoding have demonstrated outstanding performance in neuroscience~\cite{luo2023hybrid,ye2023explainable,yin2022hyperntf}. Nonetheless, the exploration of neural signal reconstruction remains nascent.


In the field of neuroscience, the reconstruction of neural signals is critically important, especially in signal completion and denoising ~\cite{yan2020reconstructing}. Electroencephalogram (EEG) and functional magnetic resonance imaging (fMRI) studies are frequently disrupted by signal loss or noise, originating from equipment defects, poor electrode contact, and physiological activities~\cite{chen2017methods,zhang2021eegdenoisenet}. Masked reconstruction pre-training methods have achieved remarkable achievements in computer vision and natural language processing, demonstrating potential in neural signal reconstruction~\cite{devlin2019bert,he2022masked}.
In natural language processing, bidirectional encoder representations from transformers (BERT) employ a masking strategy during pre-training tasks such as language modeling and next sentence prediction, enabling the model to acquire more profound and comprehensive language features~\cite{devlin2019bert}. The masking strategy involves randomly masking words in the input sequence, forcing the model to focus on all potential word combinations within the context. The masked autoencoder (MAE) model masks random patches of input images and reconstructs them to learn the latent representation of the image~\cite{he2022masked,xie2022simmim}, after the vision transformer (ViT) model addressed issues related to mask tokens and positional embeddings~\cite{dosovitskiy2021an}. The MAE has not only achieved outstanding results in image reconstruction but has also been extended to multimedia applications such as audio and video~\cite{feichtenhofer2022masked,huang2022masked}. This study proposes applying the MAE model to reconstruct resting-state and task-based fMRI, which is trained by randomly masking temporal and spatial dimensions within the fMRI data.

The brain demonstrates exceptional capabilities, notably the ability to generalize learned knowledge to new tasks and to handle multiple tasks efficiently~\cite{flesch2023continual,garner2023knowledge}. Understanding the relationships among cognitive tasks is essential for comprehending how the brain processes and coordinates cognitive functions. Relationships exist not only among cognitive tasks but also among various tasks processed by deep learning models. The effectiveness of transfer learning is influenced by the relationships between tasks. Transfer learning can enhance performance on target tasks by transferring knowledge from source tasks. The closer the relationship between the tasks, the better the results of transfer learning tend to be~\cite{pan2009survey,yosinski2014transferable}. Researchers quantify the relationships among computer vision tasks by analyzing the changes in performance of transfer learning~\cite{zamir2018taskonomy}. The cognitive taskonomy not only deepens our understanding of the relationships among tasks but also guides the selection of more relevant source tasks to improve the performance on target tasks~\cite{zamir2018taskonomy}. Compared to neural decoding classification models~\cite{qu2022transfer}, the MAE model provides a more profound approach to investigating the relationships among cognitive tasks. The MAE model effectively captures the temporal dynamic patterns and the interactions between brain regions, thereby enriching the study of cognitive task relationships. Investigating the performance changes in transfer learning via the MAE model helps to illuminate the relationships among cognitive tasks.\\


\begin{figure}[t]
\centering
\includegraphics[width=0.95\linewidth]{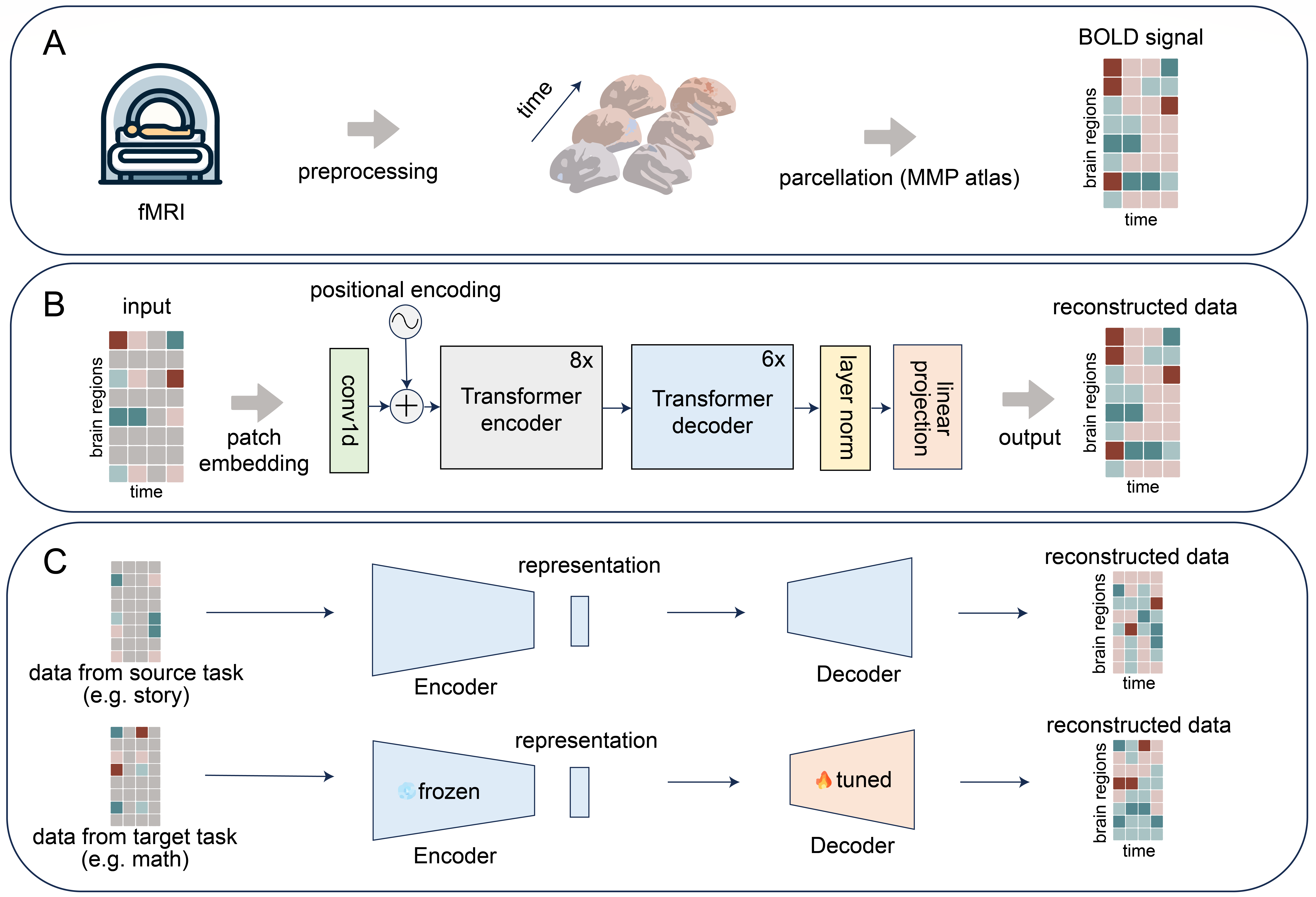}
\caption{Framework for fMRI reconstruction and transfer learning: (A) fMRI preprocessing, which involves preprocessing fMRI data and parcellating the brain using the MMP atlas. (B) The MAE model for fMRI reconstruction. Masked fMRI data are first processed through patch embedding, then passed through the transformer-based encoder and decoder, and finally linearly projected to reconstruct the fMRI signals. (C) Transfer learning is utilized to estimate the relationships among the source and target tasks. Initially, the MAE model is trained on the source task using masked fMRI data from the source task. In the second step, the MAE model, trained on the source task, is transferred to the target task. During this specific training phase, which utilizes masked fMRI data from the target task, the encoder is frozen to prevent updates, while only the decoder is tuned.}
\label{figure-1}
\end{figure}

In this study, we employ the MAE model to reconstruct fMRI signals in both resting and task states, and utilize a transfer learning framework to quantify the relationships among cognitive tasks. The contributions of this research are as follows: (1) The MAE model demonstrates robust fMRI reconstruction capabilities across individuals in resting and task states, effectively learning the temporal dynamic patterns and interactions within the brain in fMRI data. (2) Mask testing reveals the difficulty of fMRI reconstruction of different brain networks, highlighting the diversity of these networks. Furthermore, the reconstruction difficulty varies among cognitive tasks, with working memory tasks being more challenging, while language and motor tasks are comparatively easier. (3) The application of the transfer learning framework further quantifies the relationships between cognitive tasks, revealing subtask correlations within motor tasks and similarities among emotion, social, and gambling tasks. The robust cross-subject fMRI reconstruction enabled by the MAE model holds the potential for addressing common issues of signal loss in neuroimaging data collection~\cite{yan2020reconstructing}. Moreover, the task relationship analysis derived from the transfer learning framework offers guidance for selecting source tasks in neural decoding research.


\section{Method}

\subsection{HCP Dataset}


In this study, the fMRI data utilized were obtained from the public large-scale Human Connectome Project (HCP) S1200 dataset~\cite{van2012human}. The fMRI data utilized in this study comprises resting-state and task-based fMRI data from 1,000 healthy participants. These fMRI data were preprocessed using the standard HCP pipeline~\cite{glasser2013minimal}, and then parcellated into 360 brain regions based on the multi-modal parcellation (MMP) atlas~\cite{glasser2016multi}. The brain regions were divided into 7 networks according to the Yeo-7 network, including the visual network (VIS), somatomotor network (SOM), dorsal attention network (DAN), ventral attention network (VAN), limbic network (LIM), frontoparietal network (FPN), and default mode network (DMN)~\cite{yeo2011organization}.


The task-based fMRI data from the HCP comprises seven categories of cognitive tasks: working memory, motor, emotion processing, gambling, language processing, relational processing, and social cognition. These categories are subdivided into 23 subtasks, utilized by the MAE model for task-based fMRI reconstruction. 
The working memory task requires participants to memorize and recognize a series of images, divided into eight subtasks based on target categories (body, faces, places, tools) and recall steps (0-back, 2-back): 0bk body, 0bk faces, 0bk places, 0bk tools, 2bk body, 2bk faces, 2bk places, and 2bk tools. The motor task elicits simple bodily responses through visual cues, divided into five subtasks based on the body parts involved: left foot, right foot, left hand, right hand, and tongue. The emotion processing task, which assesses the ability to recognize emotions, is divided into two subtasks: fear and neutral. The gambling task examines decision-making and risk assessment through a simulated card game and includes two subtasks: win and loss. The language processing task involves auditory comprehension and language reasoning, and is divided into two subtasks: math and story. The relational processing task tests logical reasoning by comparing the shapes and textures of objects and comprises two subtasks: relation and match. The social cognition task evaluates understanding of social interactions by judging the intentions of geometric shapes' actions and includes two subtasks: mental and random.

\subsection{A Masked fMRI Modeling Framework}

Inspired by the MAE and SIMM models~\cite{he2022masked,xie2022simmim}, we propose a masked fMRI reconstruction model that incorporates masking strategies, patch embedding with position encoding, an encoder, and a decoder. The encoder of this model initially maps the masked fMRI signals to a latent representation space, followed by the decoder, which reconstructs fMRI signals based on latent representations. In this study, we utilized three masking strategies: brain masking, time masking, and a brain and time masking, as illustrated in Fig.\ref{figure-1}. In brain masking, signals from brain regions are masked throughout the entire fMRI time series. In time masking, the signal at the masked time points across all brain regions is masked. The brain and time masking  randomly mask both dimensions to learn the temporal dynamic patterns of brain activity and the interactions within the brain regions for the MAE model.

The preprocessed fMRI data were initially divided into patches and masked accordingly. Unlike MAE, which encodes only the unmasked patches, our approach, similar to simMIM~\cite{xie2022simmim}, encodes all patches using 1D convolutional neural network. After the patch embedding process, the encoded patches were given sinusoidal positional encoding. The encoder is an 8-layer transformer encoder structure that processes the input data. The decoder, which is a 6-layer transformer decoder structure, applies layer normalization and performs a linear projection to reconstruct the fMRI data. The model utilizes mean squared error (MSE) as the loss function to evaluate the difference between the reconstructed data and the original data, with optimization of model parameters occurring through backpropagation.


\begin{figure}[!t]
\centering
\includegraphics[width=0.9\linewidth]{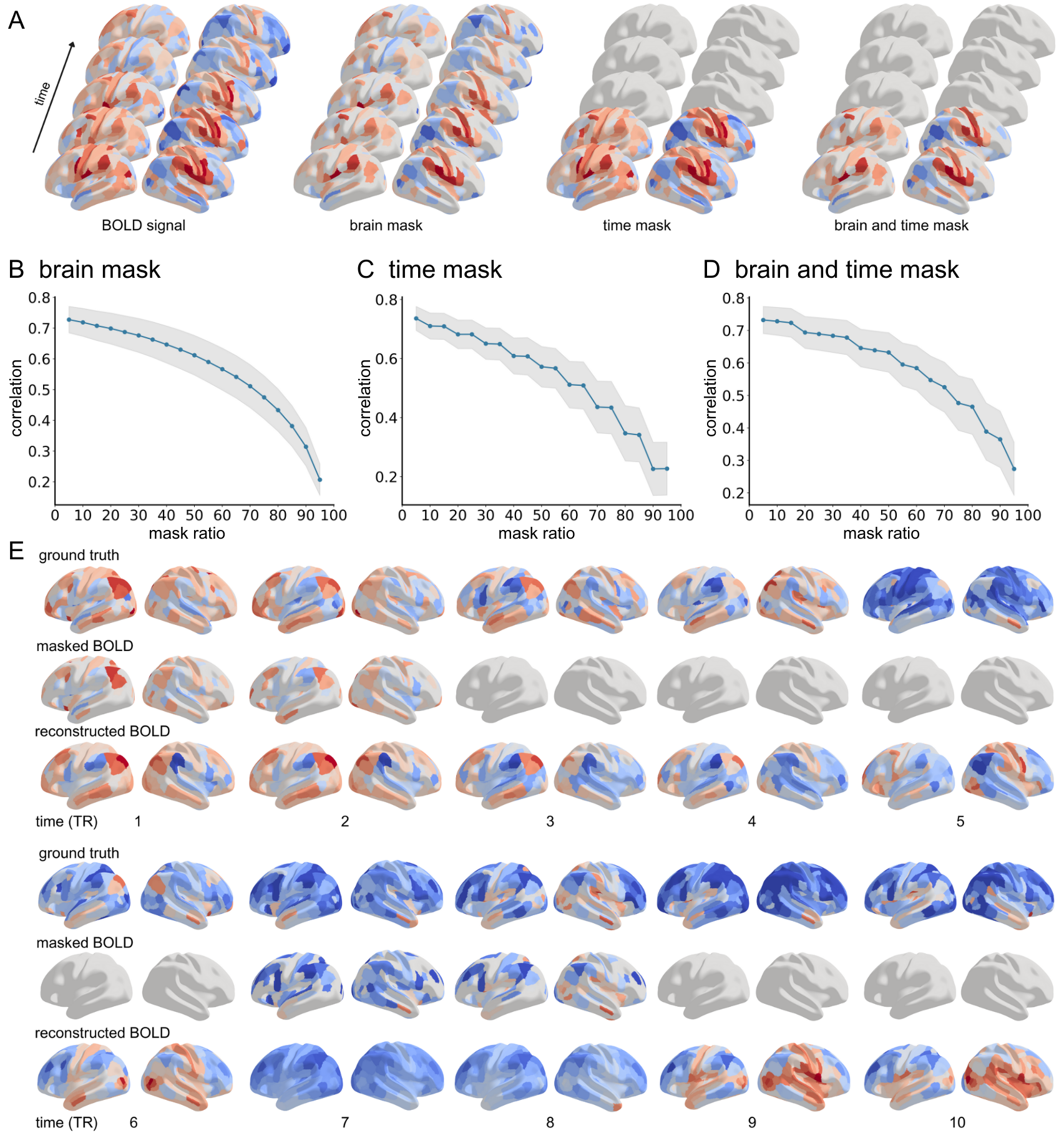}
\caption{Results of resting-state fMRI reconstruction using the MAE model. (A) Different masking strategies. Original data, brain-masked data, time-masked data, both brain and time masked data are shown from left to right. (B), (C), and (D) show the reconstruction results of the MAE model with varying mask ratios using the brain mask strategy, the time mask strategy, and the brain and time mask strategy, respectively. (E) Illustration of fMRI reconstruction by the MAE model of the brain and time mask strategy. The original signal, the masked BOLD signal, and the MAE reconstruction results are shown from top to bottom.}
\label{figure-2}
\end{figure}

\begin{figure}[t]
\centering
\includegraphics[width=0.95\linewidth]{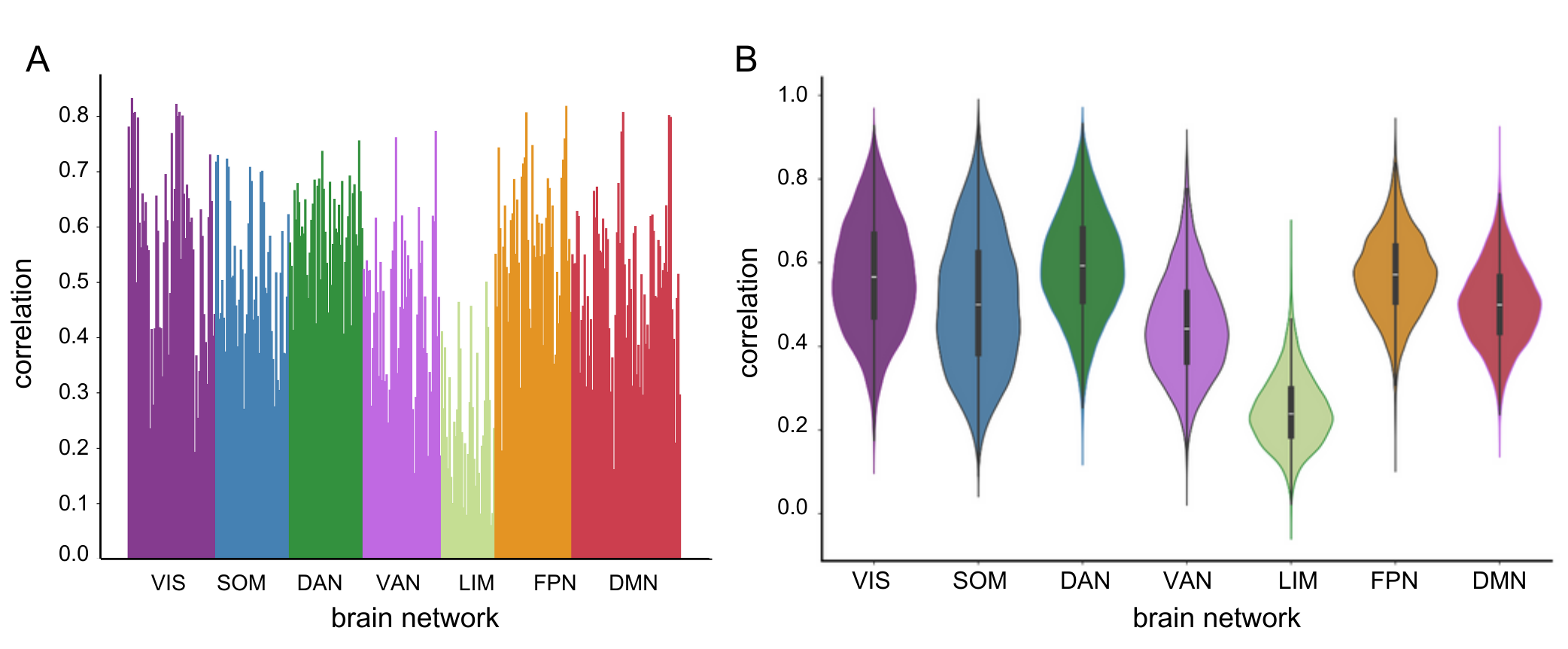}
\caption{Results of brain mask testing. (A) The MAE reconstruction results for 360 brain regions (sorted according to brain networks). (B) The MAE reconstruction results for each brain network.}
\label{figure-3}
\end{figure}

\begin{figure}[t]
\centering
\includegraphics[width=0.90\linewidth]{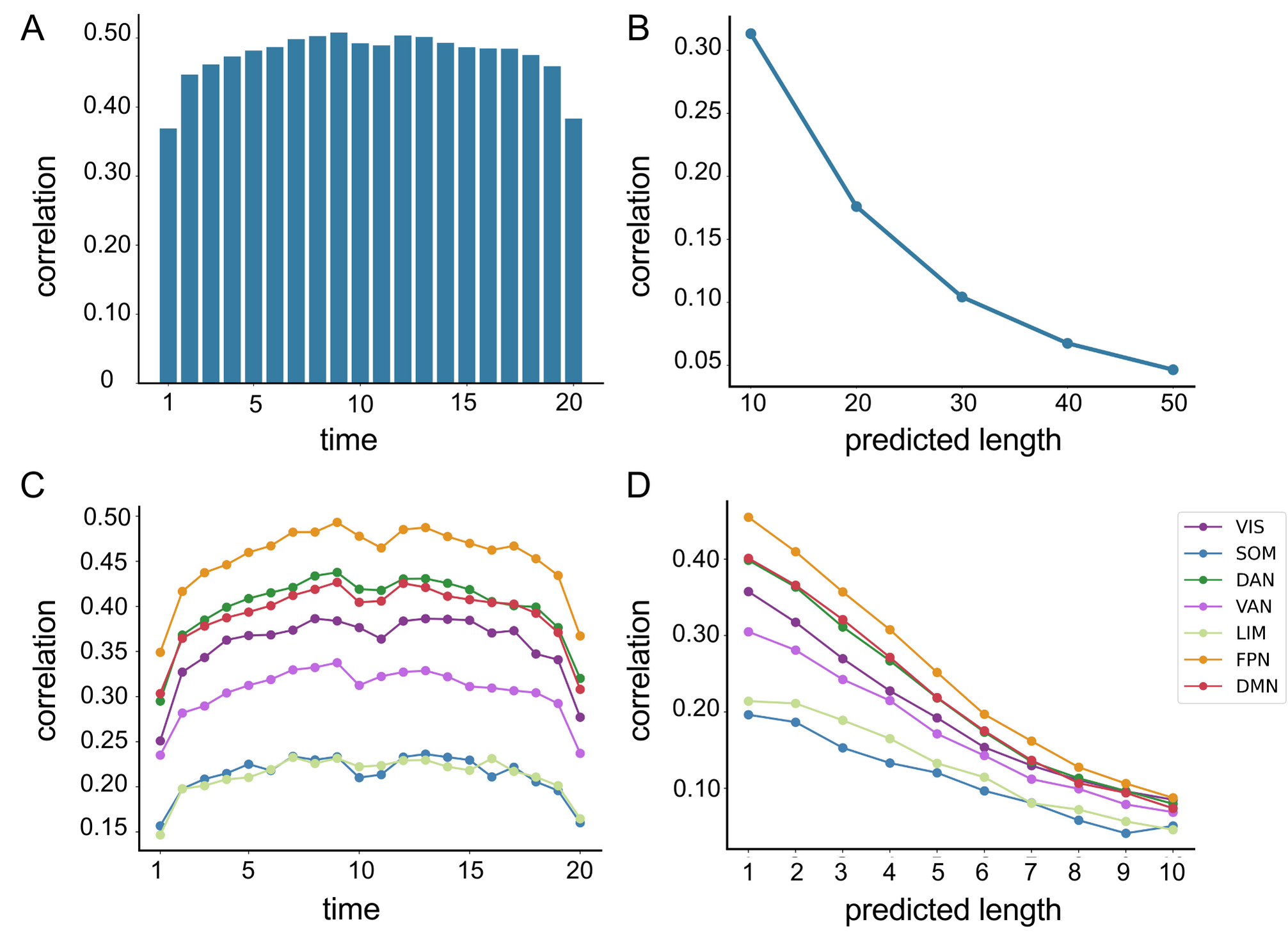}
\caption{Results of time mask testing. (A) The MAE reconstruction results for masked fMRI data, with each time frame individually masked. (B) The MAE prediction results for fMRI signals at various future time scales. (C) The MAE reconstruction results of different brain networks for masked fMRI data, with each time frame individually masked. (D) The MAE prediction results of different brain networks for fMRI signals at various future time scales.}
\label{figure-4}
\end{figure}

\subsection{Transfer learning}

The fMRI reconstruction model trained on each cognitive subtask serves as the gold standard for the source tasks. When transferring the source task to the target task, the encoder parameters of the gold standard model were fixed, and the decoder was fine-tuned using 1\% of the target task data. To construct the cognitive task relationship matrix (cognitive taskonomy), we conducted transfer experiments among the 23 cognitive tasks. Each cognitive subtask was used as a source task and transferred to the other 22 tasks, forming a $23 \times 23$ matrix. The elements $(i, j)$ represented the transfer from source task $i$ to target task $j$. The value for each matrix element was defined as the MSE of the transferred model minus the MSE of the gold standard model (trained with 80\% of the data). Finally, we applied z-score normalization to each column of the matrix, resulting in the cognitive taskonomy, as shown in Fig.\ref{figure-6}.



\subsection{Experimental setup}

In this study, we comprehensively tested the performance of fMRI reconstruction under different configurations of MAE models. The preprocessed fMRI data were divided into a 2D matrix consisting of 360 brain regions and 20 time frames, ensuring consistent time lengths for both resting-state and task-based fMRI. Specifically, we explored masking ratios of 25\%, 50\%, and 75\%, and evaluated the impact of patch sizes of 2, 5, and 10 on the fMRI reconstruction performance. Additionally, we evaluated the effect of varying the encoder and decoder with different hidden size of 512, 768, and 1024 on the reconstruction results. Based on the ablation study results (Table~\ref{table-ablation}), we selected a random masking ratio of 50\%, set the patch size in the temporal dimension to 2, and chose hidden size of 1024 for both the encoder and decoder. The fMRI data were divided into training and test sets at a 4:1 ratio based on different subjects. During training, the learning rate for the MAE model was set to 3e-5, using the AdamW optimizer, and a cosine annealing strategy was employed to adjust the learning rate.

\begin{table}[ht]
    \caption{Model ablation experiment for resting-state fMRI reconstruction. We report masking ratio (mr), correlation (corr), encoder hidden size (ehs) and decoder hidden size (dhs). If not specified, the default settings are a 50\% masking ratio, a patch size of 2, and encoder and decoder hidden sizes of 1024. Default settings are marked in \colorbox{gray!25}{gray}.}
    \begin{subtable}{.5\linewidth}
    \centering
    \caption{Masking ratio}
    \setlength{\tabcolsep}{7pt} 
    \begin{tabular}{cccc}
    \toprule   
       mr & corr & mae & mse \\ \midrule
       0.25 & 0.576 & 0.228 & 0.092 \\  
        \rowcolor{gray!25} 0.5 &\textbf{ 0.614} &\textbf{ 0.220} & \textbf{0.082} \\  
       0.75 &  0.593 & 0.221 & \textbf{0.082} \\  
    \bottomrule
    \end{tabular}
    \label{Masking ratio}
    \end{subtable}
    \begin{subtable}{.5\linewidth}
    \centering
    \caption{Patch size}
    \setlength{\tabcolsep}{7pt} 
    \begin{tabular}{cccc}
    \toprule   
       ps & corr & mae & mse \\ \midrule
       \rowcolor{gray!25} 2 & \textbf{ 0.614} & \textbf{0.220} & \textbf{0.082} \\  
       5 & 0.527 & 0.241 & 0.098 \\  
       10 & 0.438 & 0.256 & 0.112 \\  
    \bottomrule 
    \end{tabular}
    \label{Patch size}
    \end{subtable}
    \begin{subtable}{.5\linewidth}
    \centering
    \caption{Encoder hidden size}
    \setlength{\tabcolsep}{7pt} 
    \begin{tabular}{cccc}
    \toprule   
       ehs & corr & mae & mse \\ \midrule
       512 & 0.586 & 0.228 & 0.087\\  
       768 &  0.589 & 0.227 & 0.088 \\  
       \rowcolor{gray!25} 1024 & \textbf{0.614} & \textbf{0.220} & \textbf{0.082 }\\
    \bottomrule 
    \end{tabular}
    \label{Encoder hidden size}
    \end{subtable}
    \begin{subtable}{.5\linewidth}
    \centering
    \caption{Decoder hidden size}
    \setlength{\tabcolsep}{7pt} 
    \begin{tabular}{cccc}
    \toprule   
       dhs & corr & mae & mse \\ \midrule
       512 & 0.589 & 0.227 & 0.087 \\  
       768 & 0.601 & 0.224 &0.085 \\  
       \rowcolor{gray!25} 1024 & \textbf{0.614} & \textbf{0.220} & \textbf{0.082} \\
    \bottomrule 
    \end{tabular}
    \label{Decoder hidden size}
    \end{subtable}
    \label{table-ablation}
\end{table}


\section{Results}

\subsection{Reconstruction of resting-state and task-based fMRI}

To assess the impact of different masking strategies on fMRI reconstruction, we used the Pearson correlation coefficient to evaluate the reconstruction results, as shown in Fig.\ref{figure-2}. The reconstruction results of the MAE model with varying mask ratios indicate that the model can effectively learn latent features in both spatial and temporal dimensions, achieving high quality fMRI reconstruction. Although increasing the masking ratio usually leads to worse reconstruction performance, the Pearson correlation coefficient of the model reconstruction results still exceeds 0.5 at high mask ratios (\textit{e.g.}, 0.6), demonstrating the model's excellent reconstruction ability.
To comprehensively evaluate the impact of different parameter settings on performance, we conducted ablation experiments involving various masking ratios, patch sizes, and hidden sizes, as shown in Table~\ref{table-ablation}. A comparison of different patch sizes showed that smaller patch sizes resulted in lower reconstruction loss. 
Table~\ref{table-ablation} shows that encoders and decoders with larger hidden size achieve better reconstruction performance.

\begin{figure}
\centering
\includegraphics[width=0.95\linewidth]{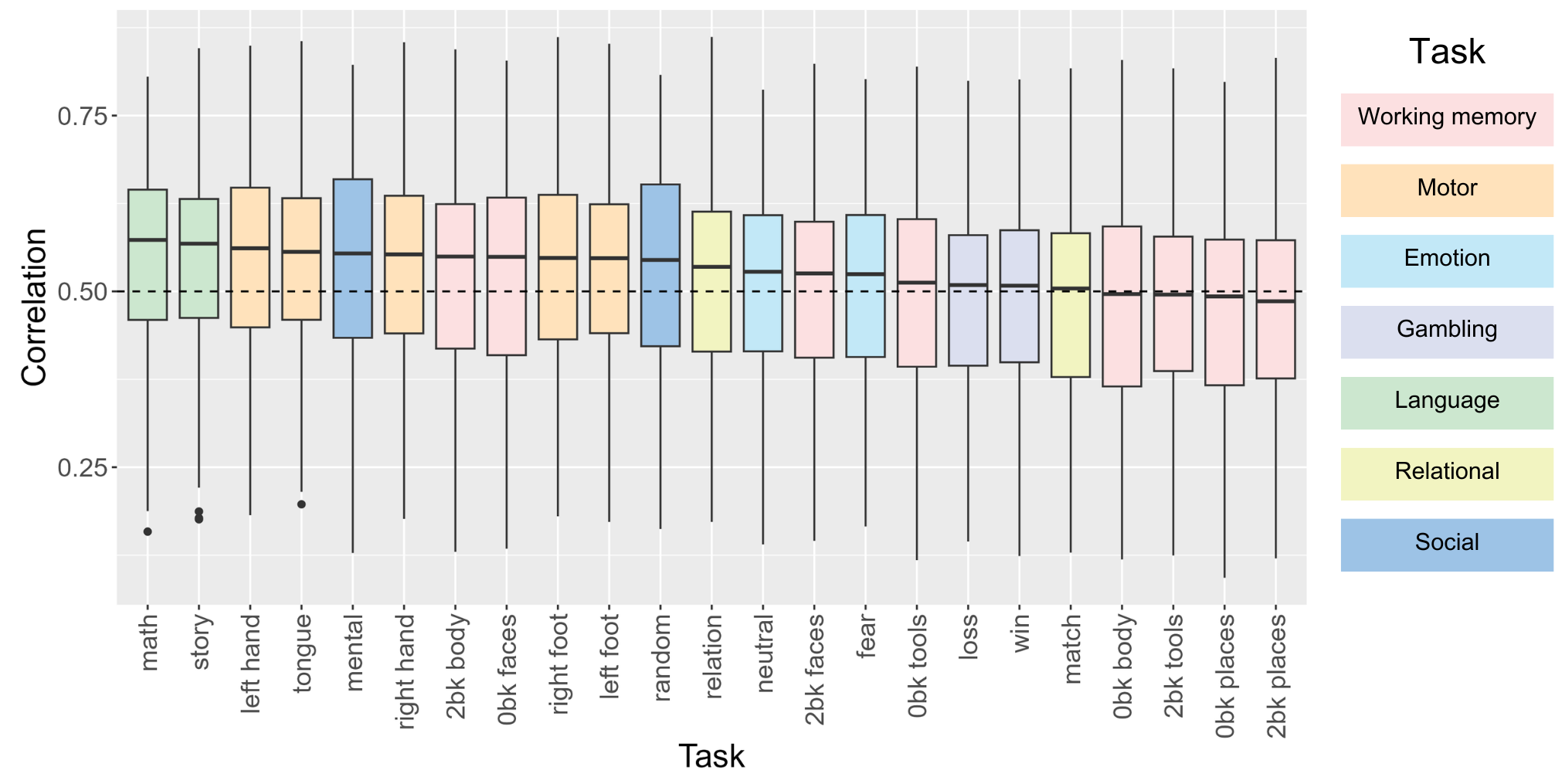}
\caption{Results of fMRI reconstruction across cognitive tasks using the MAE model. The tasks are sorted by the correlation coefficient of the reconstruction results, from highest to lowest. The colors correspond to the cognitive categories to which each subtask belongs.}
\label{figure-5}
\end{figure}

To explore the task-based fMRI reconstruction performance, MAE models were trained using fMRI data from various cognitive tasks. Fig.\ref{figure-5} depicts the reconstruction results for each task. The experimental results indicate that the Pearson correlation coefficient of fMRI reconstruction for most cognitive tasks are above 0.5, and relatively similar among subtasks within the same cognitive category. Among 23 cognitive subtasks, the reconstruction results for language tasks were the best, whereas those for working memory tasks were relatively lower. Fig.\ref{figure-5} illustrates the differing difficulties encountered in reconstructing fMRI data under various cognitive tasks, with language and motor tasks being easier to reconstruct and working memory and gambling tasks being more challenging.

\subsection{Brain mask and time mask testing}

To comprehensively analyze the differences in reconstruction difficulty among brain regions, we masked each brain region individually and calculated the Pearson correlation coefficients for the reconstruction results. Fig.\ref{figure-3} shows the results of the reconstruction of each brain region and brain networks.
The reconstruction correlations for brain regions within the LIM network have a high probability of falling below 0.4, indicating suboptimal reconstruction performance. Conversely, the median reconstruction correlations for the VIS, DAN, and FPN networks exceed 0.5, suggesting these regions more effectively utilize information from other regions for fMRI reconstruction.

In addition to analyzing the differences in reconstruction among various brain regions, the properties of MAE in fMRI reconstruction along the temporal dimension were further investigated. Fig.\ref{figure-4}A displays the Pearson correlation coefficients for the reconstructed results of fMRI data, with each time frame individually masked. The results indicate that signals in the middle of the time series are relatively easier to reconstruct, while those at the beginning and end of the series pose greater challenges. Fig.\ref{figure-4}B illustrates the correlation between the length of the prediction interval and the reconstruction performance. In this analysis, we use a fixed set of 20 frames of data to predict future signals of different lengths. The results demonstrate that as the prediction interval increases, the reconstruction performance gradually declines. Fig.\ref{figure-4}C and Fig.\ref{figure-4}D show that brain networks with strong signal reconstruction capabilities at various time points also perform well in predicting signals of varying lengths. For example, LIM and SOM networks do not have a strong ability to reconstruct signals at different points in time, nor can they effectively predict future signals.


\begin{figure}
\centering
\includegraphics[width=0.65\linewidth]{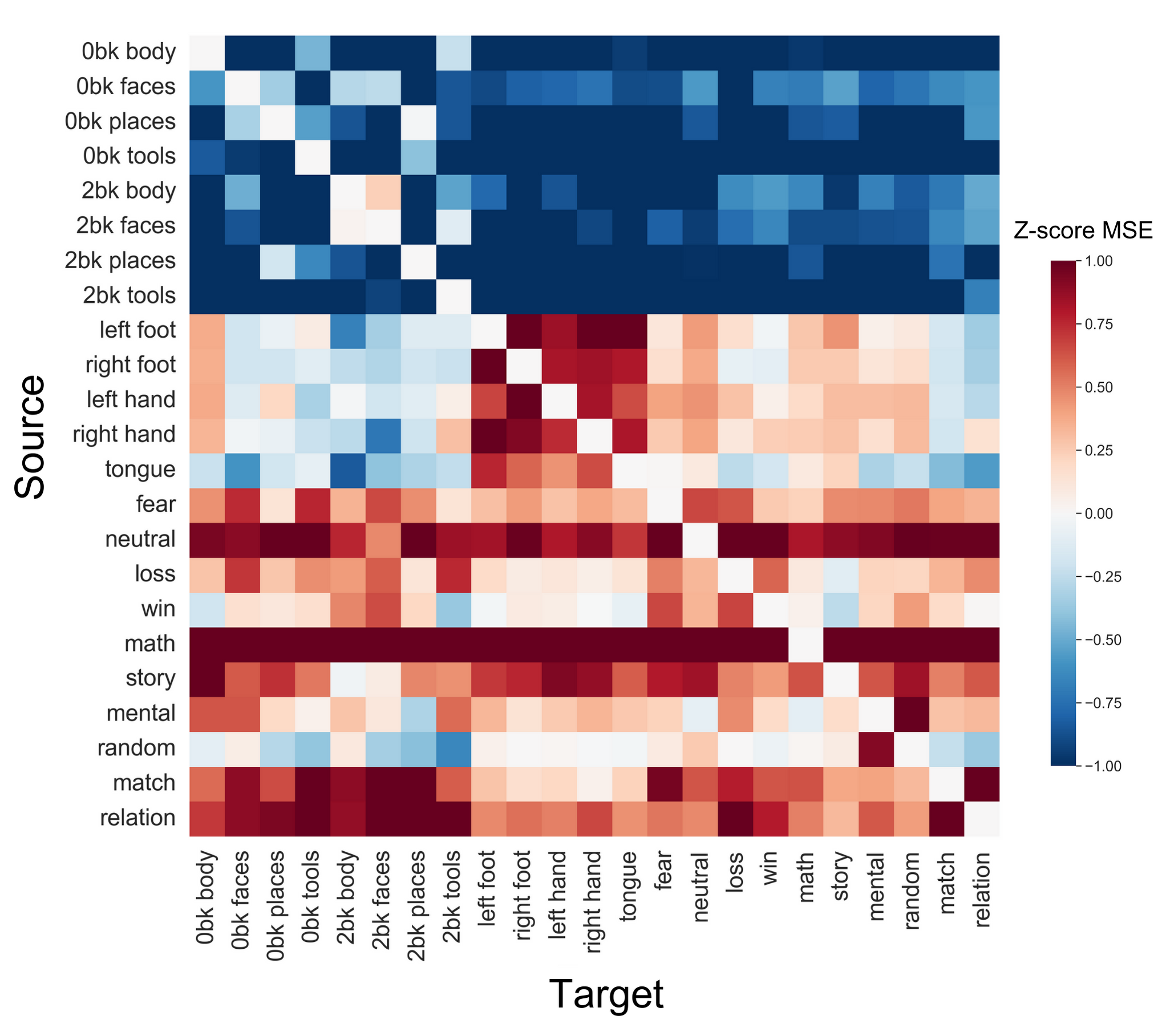}
\caption{Cognitive taskonomy derived from transfer learning, demonstrates the similarity among 23 cognitive tasks. These values represent changes in the reconstruction result when transferring from source tasks to target tasks. Positive values indicate that the source task improves the reconstruction results of the target task, and negative values indicate the opposite.
}
\label{figure-6}
\end{figure}

\subsection{Relationships among cognitive tasks}

This study employs a masked fMRI modeling framework with transfer learning to quantify the relationships between cognitive tasks. As illustrated in Fig.\ref{figure-1}C, the transfer learning process comprises two steps. Initially, the MAE model is trained on masked fMRI data from the source task. After training, the MAE model is transferred to the target task to evaluate the transfer performance. During fine-tuning on the target task, the parameters of the encoder are fixed, while only the parameters of the decoder are updated using 1\% of the target task data. Finally, the relationship between tasks is determined by the transfer performance. Better transfer performance indicates a closer relationship between the source and target tasks, whereas poorer performance suggests a more distant relationship.

Fig.\ref{figure-6} depicts the transfer results among the 23 cognitive tasks, reflecting the relationships between these tasks. Tasks within the same cognitive category exhibit closer relationships. For instance, the five subtasks within the motor category demonstrate strong transferability among each other, particularly between the left foot and right foot, and between the left hand and right hand. Furthermore, transferability varies across different cognitive tasks. For example, tasks such as math and neutral tasks exhibit effective transferability to other tasks, whereas working memory tasks face challenges in transferring to other tasks.

\begin{figure}[t]
\centering
\includegraphics[width=0.98\linewidth]{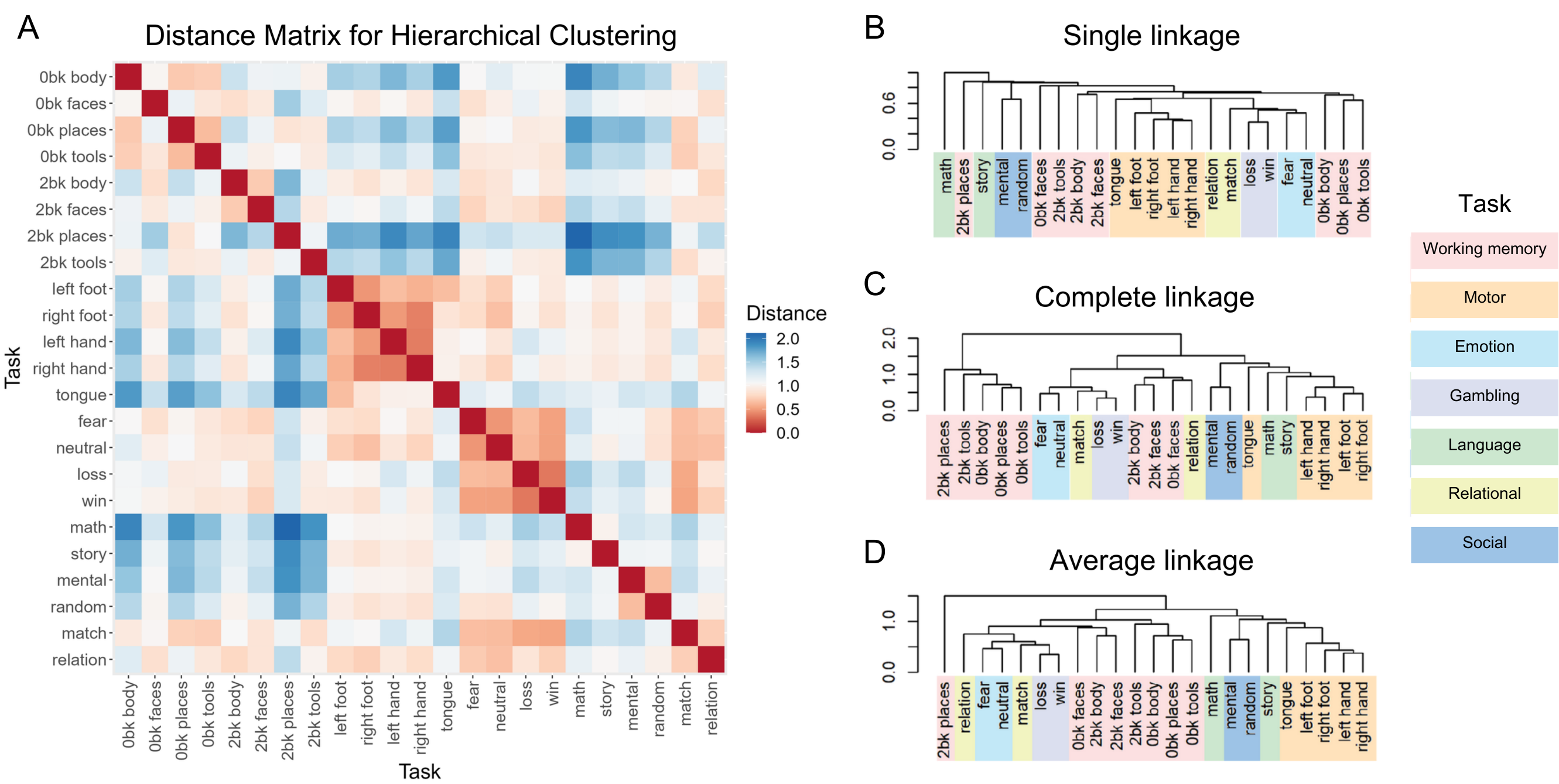}
\caption{Hierarchical clustering results of cognitive tasks. (A) The distance matrix among 23 cognitive tasks, calculated using Euclidean distances between pairs based on the reconstruction results of brain regions for each task. (B), (C), and (D) show the results of hierarchical clustering performed with single linkage, complete linkage, and average linkage methods, respectively. The colors on the right indicate the cognitive categories to which each subtask belongs.}
\label{figure-7}
\end{figure}

Fig.\ref{figure-7} presents the hierarchical clustering results of cognitive tasks based on the Euclidean distance matrix calculated from reconstruction results between tasks. The results demonstrate that tasks within the same cognitive category are more closely clustered. For instance, the five subtasks in the motor category are closely clustered, and the random and mental tasks within the social category are in close proximity. Notably, within the working memory category, 0-back places are closer to 0-back tools, and 2-back faces are closer to 2-back bodies. This indicates a stronger relationship between places and tools, as well as between faces and bodies. Fig.\ref{figure-7}B, Fig.\ref{figure-7}C, and Fig.\ref{figure-7}D employ various hierarchical clustering strategies, which generally group tasks within the same cognitive category together, further illustrating the close relationships among similar cognitive tasks.


\section{Discussion}

In the brain masking experiments, it was observed that the reconstruction results for the VIS, DAN, and FPN networks were better. The redundancy and consistency of information within brain networks may account for these differences~\cite{luppi2022synergistic,luppi2024information}.
Information from other related regions that are more consistent with the masked brain region will assist in its reconstruction.
Furthermore, some regions within these networks may demonstrate functional specialization in processing specific tasks, indicating a functional similarity among these regions~\cite{zeki1991direct}. This enables related regions to more effectively reconstruct missing information during reconstruction. Conversely, it was found that the LIM network presents more challenges in reconstruction compared to other brain networks. This is potentially due to the regions within the limbic system performing multiple functions~\cite{rolls2019cingulate}. This multifunctionality amplifies the complexity of neural activity patterns, making it more difficult for the MAE model to capture the dynamic patterns of these brain regions compared to other brain networks. In the time masking experiments, we observed that reconstruction performance was relatively poor at the beginning and end of the sequence. In tests predicting fMRI signals across various time lengths, it was found that the longer the prediction time, the poorer the performance. This can be attributed to the inherent complexity and spontaneous activation of resting-state fMRI signals~\cite{power2014studying,xing2023individual}. Brain activity is characterized by highly complex and dynamic features, influenced by a variety of internal and external factors. As the prediction length increases, this complexity and dynamism render accurate predictions increasingly challenging.

In this study, reconstruction experiments on task-based fMRI signals reveal that language and motor tasks were easier to reconstruct, while working memory tasks were more difficult. This difference is probably related to the inherent complexity of these tasks. Firstly, the brain regions activated by motor and language tasks tend to be localized~\cite{barch2013function}. In contrast, working memory tasks involve multiple cognitive processes that require coordination of the brain~\cite{collette2002brain,dong2015individual}, which increases the complexity of reconstruction.
Furthermore, individual variability across different cognitive tasks may also contribute to the observed differences in reconstruction difficulty~\cite{baldassarre2012individual,xing2023individual}. Working memory tasks often exhibit greater variability between individuals, making it more difficult for the model to reconstruct fMRI signals across different subjects~\cite{dong2015individual}.

To explore the relationships between cognitive tasks, transfer learning experiments were conducted across cognitive tasks. Better transfer performance between cognitive tasks indicates a closer relationship among them. Fig.\ref{figure-6} reveals that subtasks within the same category tend to have better transfer results. For instance, effective transfer performance is observed among the five motor subtasks. These subtasks primarily engage the primary motor cortex and premotor cortex, where the activation patterns and representations are more similar, thereby facilitating generalization between these tasks during transfer learning~\cite{gordon2023somato}. The distance matrix in Fig.\ref{figure-7} illustrates that the five motor subtasks cluster closely, reflecting the similarity among these subtasks. Additionally, differences exist in transfer ability between different cognitive tasks, with not all tasks within the same category demonstrating good transfer performance. The transfer performance of working memory tasks is poor, potentially due to the inconsistent brain regions and dynamic patterns, such as the different task difficulties with varying n-back~\cite{dong2015individual}. Similar to findings in neural decoding classification models~\cite{qu2022transfer}, fMRI reconstruction results also reveal a high similarity between emotion, social, and gambling tasks. This similarity may arise from the involvement of complex decision-making processes, which activate similar brain regions such as the prefrontal cortex~\cite{barch2013function,olsson2008role}. Overall, the transfer performance between cognitive tasks can elucidate the relationships among them. 

This study demonstrates that the MAE model can robustly reconstruct fMRI signals across subjects. This highlights the potential of using MAE models to address the common issue of signal loss in applications~\cite{yan2020reconstructing}. The analysis of reconstruction difficulty across brain regions provides a reliability assessment, enabling a more cautious interpretation of the results in brain regions that are challenging to reconstruct. Future research could extend to exploring the application of MAE models to other modalities of neural signals, such as EEG reconstruction. Furthermore, the cognitive taskonomy provides a more reliable guidance for selecting source tasks in neural decoding studies. This relationship matrix can assist researchers in more effectively utilizing source task data, thereby improving decoding performance in target tasks.





\section{Conclusion}

In this study, we employ the MAE model to reconstruct both resting-state and task-based fMRI signals and utilize a transfer learning framework to quantify the relationships between cognitive tasks. The MAE model effectively captures the temporal dynamics and interactions among brain regions, thereby enabling robust reconstruction of fMRI signals. By utilizing the transfer learning framework, we further quantify the relationships between cognitive tasks, revealing subtask correlations within motor tasks and identifying similarities among emotion, social, and gambling tasks. This study not only demonstrates the potential of the MAE model to reconstruct lost signals but also provides guidance for selecting source tasks in neural decoding studies.


\section{Acknowledgments}
This work was funded in part by the National Key R\&D Program of China (2021YFF1200804), Shenzhen Science and Technology Innovation Committee (2022410129, KCXFZ20201221173400001).

\bibliographystyle{splncs04}
\bibliography{ref}
\end{document}